\documentstyle[preprint,aps]{revtex}
\begin{document}
\draft

\preprint{gr-qc/9309026}

\title{Black Hole Entropy \\ and the \\ Dimensional Continuation
of the Gauss-Bonnet Theorem}

\author{M\'aximo Ba\~nados$^1$, Claudio Teitelboim$^{1,2}$
and Jorge Zanelli$^{1,3}$}

\address{~$^{1}$Centro de Estudios Cient\'{\i}ficos de Santiago,
Casilla 16443, Santiago 9, Chile.  \\
{}~$^{2}$ Institute for Advanced Study, Olden Lane, Princeton,
New Jersey 08540.  \\
{}~$^{3}$ Departamento de F\'{\i}sica, Facultad de Ciencias,
Universidad de Chile, \\ Casilla 653, Santiago, Chile. }

\maketitle

\begin{abstract}
\end{abstract}


\narrowtext

The purpose of this note is to point out that the black-hole
entropy may be derived in a simple manner if one regards the
Euclidean Hilbert action for the gravitational field as the
dimensional continuation of a topological invariant, the Euler
class. Then, the dimensional continuation of the Gauss--Bonnet
theorem shows that the entropy itself is the continuation of the
Euler class of a small disk centered at the horizon.

The Euclidean spacetimes admitted in the action principle will
have the topology $\Re^2 \times {\cal S}^{d-2}$. We allow for
cusps at any point in $\Re^2$.  The deficit angle of a cusp
at a given point turns out to be canonically conjugate to the
area of the ${\cal S}^{d-2}$ at that point.  In particular, the
entropy is canonically conjugate to the deficit angle at the
horizon. The condition for zero deficit angle emerges from
extremizing the action with respect to the area of the ${\cal
S}^{d-2}$.  Thus, the black hole temperature is determined, in
the semiclassical approximation, by extremizing with respect to
the area of the ${\cal S}^{d-2}$ at the horizon (summing over
all horizons in the path integral).

This result applies also to the natural generalization of the
Hilbert action to higher spacetime dimensions, the
Lovelock action \cite{Lovelock}. This action, which keeps the
field equations for the metric of second order and hence does
not change the degrees of freedom, can also be
understood in terms of dimensional continuation
\cite{Zumino,Teitelboim-Zanelli}. For a spacetime of dimension
$d$, the generalized action contains the dimensionally continued
Euler classes of all even dimensions $2p<d$. Each such term
gives rise to an entropy \cite{Jacobson-Myers} that is
proportional to a dimensional continuation of the Euler class of
dimension $2p-2$.  Thus, the Hilbert action with a cosmological
constant may be thought of as coming from dimensions $2p=2$ and
$2p=0$, respectively. The entropy comes then from $2p=0$. (We
define the Euler class for a space of dimension zero as unity
and that for a space of negative dimension as zero.)

We will first recall the Gauss-Bonnet theorem and bring out its
relationship with the Hilbert action for the gravitational field.
The extension to the Lovelock theory will be indicated at the end.

If one considers a two dimensional manifold $M$ with
boundary $\partial M$, the Gauss--Bonnet theorem reads

\begin{equation}
\frac{1}{2}\int_M \sqrt{g} g^{\mu \nu} R^{\alpha}_{\;\; \mu
\alpha \nu}d^2x - \int_{\partial M} \sqrt{g}K d^1x =
2\pi\chi(M).
\label{1}
\end{equation}

The integer $\chi(M)$ on the right hand side of (\ref{1}) is the
Euler number of $M$ and depends solely on its topology. One has
$\chi=1$ for a disk and $\chi =0$ for an annulus. We will refer
to the sum of integrals appearing on the left side of (\ref{1})
as the Euler class of $M$. The Gauss-Bonnet theorem then says
that the Euler class of $M$ is equal to $2\pi$ times its Euler
number.

If one varies the integral over $M$ in (\ref{1}) one finds, by
virtue of the Bianchi identity, that the piece coming from the
variation of the Riemann tensor yields a surface term. This
surface term exactly cancels the variation of the surface
integral appearing in the Euler class. On the other hand,
because of the special algebraic properties of the Riemann
tensor in two spacetime dimensions, the contribution of the
variation of $\sqrt{g}g^{\mu \nu}$ is identically zero.  This
is a poor man's way to put into evidence that the Euler class is
``a topological invariant", the real work is to show that the
actual value of the sum of integrals is $2\pi \chi$.

Now, the Hilbert action for the gravitational field in $d$
Euclidean spacetime dimensions may be written as

\begin{equation}
I_H = \frac{1}{2}\int_M \sqrt{g} g^{\mu \nu}
R^{\alpha}_{\;\;\mu \alpha \nu}d^dx - \int_{\partial M} \sqrt{g}
K d^{d-1}x .
\label{2}
\end{equation}
[One integrates $\exp(+I)$ in the Euclidean path integral.]

This action has the same form as the Euler class of two
dimensions, with the change that now the integrals, and the
geometric expressions appearing in them, refer to a spacetime of
dimension $d>2$. For this reason, one says that the Hilbert
action is the dimensional continuation of the Euler class of two
dimensions. After dimensional continuation, the Euler class
ceases to be a topological invariant. While it is still true
that the variation of the Riemann tensor in (\ref{2}) yields a
surface term, this surface term no longer cancels the variation
of the integral of the extrinsic curvature. Rather, the sum of
the two variations vanishes only when the intrinsic geometry of
the boundary is held fixed. Moreover, the contribution to the
variation coming from $\sqrt{g}g^{\mu \nu}$ gives the
Einstein tensor, which is no longer identically zero, and hence
the demand that it vanishes is not empty but gives the Einstein
equations.

There is another action, which differs from the $I_H$ by a
boundary term. It is the canonical action

\begin{equation}
I_C = \int ( \pi^{ij} \dot{g}_{ij} - N {\cal H} -
N^{i}{\cal H}_i).
\label{3}
\end{equation}

When one studies black holes $I_C$ has a significant advantage
over the Hilbert action. It vanishes on the black hole due to
the constraint equations ${\cal H} =0= {\cal H}_i$ and the time
independence of the spatial metric.  The black hole entropy and
its relation with the Gauss-Bonnet theorem will arise through
the difference between the Hilbert and the canonical actions.

In the Euclidean formalism for black holes, it is useful to
introduce in the $\Re^2$ factor of $\Re^2 \times {\cal
S}^{d-2}$, a polar system of coordinates. The reason is that the
black hole will have a Killing vector field --the Killing time--
whose orbits are circles centered at the horizon. But, it should
be stressed that the discussion that follows is valid for a
system of polar coordinates centered anywhere in $\Re^2$. Indeed
the Killing vector exists only on the extremum and not for a
generic spacetime admited in the action principle.

Take now a polar angle in $\Re^2$ as the time variable in a
Hamiltonian analysis. An initial surface of time $t_1$ and a
final surface of time $t_2$ will meet at the origin, which is a
fixed point of the time vector field. There is nothing wrong
with the two surfaces intersecting. The hamiltonian formalism
can handle that.

Consider now in $\Re^2$ a small disk $D_{\epsilon}$ of radius
$\epsilon$ around the origin. The portion of spacetime between
$t_1$ and $t_2$ that remains inside the small disk is a triangle
$\triangle_{\epsilon}$, of angular opening $t_2-t_1$, to each
point of which an ${\cal S}^{d-2}$ sphere is attached. The
action (\ref{3}) for $\triangle_{\epsilon}\times {\cal S}^{d-2}
$ tends to zero if one lets $\epsilon \rightarrow 0$ because the
integrand is smooth. On the other hand, the covariant action
(\ref{2}) tends to a non-zero limit --which will be exhibited in
Eq. (\ref{8}) below. Thus we have

\begin{mathletters}
\label{actions}
\begin{equation}
\lim_{\epsilon \rightarrow 0} I_C [\triangle_{\epsilon}\times
{\cal S}^{d-2}]  = 0 ,
\label{4'}
\end{equation}
\begin{equation}
\lim_{\epsilon \rightarrow 0} I_H [\triangle_{\epsilon}\times
{\cal S}^{d-2}]  \neq 0.
\label{5}
\end{equation}
\end{mathletters}

Next, recall that the very derivation of the canonical action
(see for example \cite{Teitelboim-Zanelli}), shows that for
surfaces of constant $t$, the inclusion of the extrinsic
curvature term precisely turns (\ref{2}) into (\ref{3}).
Combining this fact with the preceding remarks we see that for
the whole region between $t_1$ and $t_2$ we have

\begin{eqnarray}
I_H &=& \lim_{\epsilon \rightarrow 0} I_H[\triangle_{\epsilon}\times
{\cal S}^{d-2}] + I_C + B_{\infty}.
\label{6}
\end{eqnarray}
Here $B_{\infty}$, which needs not be explicitly written, stands
for a surface term over a large circle in $\Re^2 \times {\cal S}^{d-2}$.
Indeed, as stated above, the Hilbert action (\ref{2}) needs the
intrinsic geometry of the entire boundary to be fixed. On the
other hand for the Hamiltonian action (\ref{3}) one must fix at
$t_1$ and $t_2$ the intrinsic geometries of those boundaries
--including their intersection at the origin-- and, at infinity,
the mass $M$ and angular momentum $J$ --with a precise rate of
fall off for the fields [see, for example
\cite{Regge-Teitelboim}]. If instead of $M$, one fixes its
conjugate,  the asymptotic Killing time difference
$t_2-t_1=\beta$, while still keeping $J$ fixed, one must
substract $\beta M$ from (\ref{3}). Thus, if we drop
$B_{\infty}$, we obtain the improved covariant action,

\begin{equation}
I = \lim_{\epsilon \rightarrow 0} I_H[\triangle_{\epsilon} \times
{\cal S}^{d-2}] + I_C ,
\label{7}
\end{equation}
which is suited for fixing the intrinsic geometries of the
surfaces at $t_1$ and $t_2$, and $M$ and $J$ at infinity. The
action (\ref{7}) differs from expression (\ref{6}) only by a
local surface term at infinity due to the different boundary
condition there, and it is therefore as covariant as (\ref{2}).
Furthermore, (\ref{7}) is finite on the black hole and thus it
is ``already regularized''. [The Hilbert action (\ref{2}) is
infinite on the black hole because $B_{\infty}$ diverges.]

A short analysis reveals that the first term in (\ref{7})
factorizes into the product of the Euler class (\ref{1}) for
$\triangle_{\epsilon}$ and the area of the ${\cal S}^{d-2}$ at
the origin. But, the triangle $\triangle_{\epsilon}$ is
topologically a disk and hence its Euler class is equal to
$2\pi$. Thus one finds

\begin{equation}
\lim_{\epsilon \rightarrow 0} I_H [\triangle_{\epsilon} \times
{\cal S}^{d-2}] = 2\pi \times (\mbox{area of }{\cal S}^{d-2} )_{origin}.
\label{8}
\end{equation}

It is of interest to allow in (\ref{7}) for a ``cusp of deficit angle
$\alpha$'' at the origin of $\Re^2$. This means that the value
of the two-dimensional integral in the Euler class (\ref{1}) is
equal to $\alpha$, whereas the line integral over the boundary
has the value $2\pi - \alpha$. The full action (\ref{7}) depends
on $\alpha$. This is most directly seen by recalling that -as
stated in (\ref{6})- the action (\ref{7}) differs from the
Hilbert action (\ref{2}) by a local boundary term at infinity. As a
consequence, if the geometry of the ${\cal S}^{d-2}$ at the cusp is
varied, while keeping the rest of the configuration unaltered,
one finds that the action changes by

\begin{equation}
\delta I = \alpha \delta(\mbox{area of ${\cal S}^{d-2}$ at cusp})
\label{9}
\end{equation}

Equation (\ref{9}) shows that the deficit angle --which is a
property of the intrinsic Riemannian geometry of $\Re^2$--, is
canonically conjugate to the area of the ${\cal S}^{d-2}$
attached to that point --an extrinsic property.

Observe that one could incorrectly believe, due to (\ref{8}), that
the action (\ref{7}) (and hence its variation) is independent of
the deficit angle $\alpha$. What happens is that there is a
boundary term in the variation of the canonical action, coming
from space derivatives in ${\cal H}$, which cancels the
variation of the surface term in the Euler class
\cite{Brown-York} leaving (\ref{9}) as the net change.

As shown by (\ref{8}), the actions (\ref{3}) and (\ref{7})
differ by a contact transformation which depends only on (part
of) the common boundary data for each action. Thus, both actions
correctly yield Einstein's equations and on this basis they are
equally good. However, one wants to do more, one wants an
action that can also be used to evaluate the partition function.

In the semiclassical approximation, the partition function is
equal to the exponential of the classical action for a ``closed
Euclidean history". In the present case the closed Euclidean
history is obtained by making the surface $t_2$ come around a
full turn and coincide with the surface $t_1$. The triangle
$\triangle_{\epsilon}$ becomes the whole disk $D_{\epsilon}$. One
then extremizes with respect to the geometry of spacetime
keeping the asymptotic data $M$ and $J$ fixed.  For
this problem, the improved action(\ref{7}) and the canonical
action (\ref{6}) are not equivalent.

The black hole will be an extremum for the covariant action
(\ref{7}), because the demand that the variation (\ref{9})
vanishes yields $\alpha=0$ at all points, which is the condition
for the manifold to be metrically smooth.  This is a property
that the Euclidean black hole indeed posseses, since the empty
space Einstein equations are obeyed everywhere.  On the other
hand, the demand that the canonical action should have an
extremum with respect to variations of the area of the ${\cal
S}^{d-2}$, would yield $\alpha = 2\pi$ at the origin, which
would introduce a sort of source there.

Thus, adding the Hilbert action for a small disk around the
origin to the canonical action restores covariance without
introducing sources. This addition ensures that the fixed point
can be located anywhere. This must be so since the manifold has
only one boundary, that at infinity.

Note that the need for the contribution of $D_{\epsilon} \times
{\cal S}^{d-2}$ in the action for the partition function,
implies that the contribution (\ref{8}) of $\triangle_{\epsilon}
\times {\cal S}^{d-2}$ must already be
included in the action for the transition amplitude from $t_1$
to $t_2$. This is because the partition function is the trace of
the transition amplitude \cite{Carlip}.

Consider now the value of the action on the extremum. Then it is
convenient to take the polar angle to be the Killing time, for
-in that case- the spatial geometry $g_{ij}$ is time
independent.  Furthermore, since the Hamiltonian contraints
${\cal H}={\cal H}_i=0$ hold on the extremum, the value of the
improved action (\ref{7}) for the black hole is just the
contribution of the disk at the horizon.

Since in (\ref{7}) $M$ and $J$ are fixed, which corresponds to
the microcanonical ensemble, we learn that the entropy is given by

\begin{equation}
S =   2\pi \times (\mbox{area of ${\cal S}^{d-2}$})_{horizon}.
\label{10}
\end{equation}

This is the standard expression for the black hole entropy in
Einstein's theory. Note that the overall factor in front of
the area, usually quoted as one fourth in units where Newton's
constant is unity, is really the Euler class of the two-dimensional
disk.

The preceding analysis goes through step by step for the
Lovelock theory \cite{Lovelock}. The analog of the Hilbert
action given by (\ref{2}) is

\begin{equation}
I_{L} = \sum_{2p<d} \frac{\alpha_p}{2^{2p} \, p! } (I_{L}^p +
B^p) ,
\label{10'}
\end{equation}
with

\begin{equation}
I_{L}^p=  \int_M\sqrt{g} \delta_{[\alpha_1 \cdots
\alpha_{2p}]}^{[\beta_1 \cdots \beta_{2p}]} R^{\alpha_1
\alpha_2}_{\beta_1 \beta_2} \cdots R^{\alpha_{2p-1}
\alpha_{2p}}_{\beta_{2p-1} \beta_{2p}}d^dx.
\label{11}
\end{equation}
[Here the totally antisymmetrized Kronecker symbol is normalized
so that it takes the values 0, $\pm 1$].

The boundary term $B^p$ is the generalization of the integrated
trace of the extrinsic curvature in (\ref{2}). It is given by

\begin{equation}
B^p = \frac{-2}{d-2p} \int_{\partial M} d^{d-1}x
g_{ij}\pi^{ij}_{(p)}.
\label{12}
\end{equation}
Here $\pi^{ij}_{(p)}$ is the contribution of (\ref{11}) to the
momentum canonically conjugate to the metric $g_{ij}$ of
$\partial M$. It may be expressed as a function of the intrinsic
and extrinsic curvatures of the boundary
\cite{Teitelboim-Zanelli}.

For Euclidean black holes in $d$-spacetime dimensions, again with
topology $\Re^2 \times {\cal S}^{d-2}$ \cite{BTZ2}, the action
(\ref{7}) now reads

\begin{eqnarray}
I &=& \lim_{\epsilon \rightarrow 0} I_{L} [D_{\epsilon} \times
{\cal S}^{d-2}]  + I_C,
\label{13}
\end{eqnarray}
and the entropy becomes

\begin{equation}
S = \lim_{\epsilon \rightarrow 0} I_{L} [D_{\epsilon} \times
{\cal S}^{d-2}].
\label{13'}
\end{equation}
The limit (\ref{13'}) factorizes into the Euler class of the
disk, equal to $2\pi$, and a sum of dimensional continuations to
${\cal S}^{d-2}$ of the Euler classes of all even dimensions below $d-2$,

\begin{equation}
S = 2\pi \times \sum_{2p<d} \frac{\alpha_p}{2^{2(p-1)}[2(p-1)]!} S^{p-1}
\label{14}
\end{equation}
with

\begin{equation}
S^p = \int \sqrt{g}\delta_{[\alpha_1 \cdots \alpha_{2p}]}^{[\beta_1
\cdots \beta_{2p}]} R^{\alpha_1 \alpha_2}_{\;\;\;\beta_1
\beta_2}\cdots R^{\alpha_{2p-1} \alpha_{2p}}_{\;\;\;
\beta_{2p-1} \beta_{2p}} d^{d-2}x,
\label{14'}
\end{equation}
where the integral is taken over the $(d-2)$-sphere at the horizon.

The Hilbert action corresponds to $2p=2$ and the corresponding
entropy is $2\pi$ times the area. The cosmological constant term
corresponds to $2p=0$ and gives no contribution to the entropy.
Expression (\ref{14}) was first given in
\cite{Jacobson-Myers}.\\

The authors are grateful to Ted Jacobson and Robert Myers for
urging them to prepare this analysis for publication.
Appreciation is also extended to Steven Carlip and Marc Henneaux
for many rewarding discussions. This work was partially
supported by grants 0862/91, and 193.0910/93 from FONDECYT
(Chile), by a European Communities research contract, by
institutional support to the Centro de Estudios Cient\'{\i}ficos
de Santiago provided by SAREC (Sweden) and a group of chilean
private companies (COPEC, CMPC, ENERSIS).


\begin{references}
\bibitem{Carlip} Note that the term which must be added to the
canonical action does not refer to any history in time $t$. This
is a deep property of the black-hole.  Steven Carlip (private
communication) describes it by saying that the action of the
small triangle refers to ``intrinsically time independent
degrees of freedom''.
\end{references}
\end{document}